\documentclass[12pt]{emulateapj}

\newcommand{\dd}{{\rm d}}

\usepackage{amssymb}
\usepackage{amsmath,textcomp,array}
\usepackage{multirow}

\submitted{Draft version}
\shorttitle{Non-Gaussianity from weak lensing}
\shortauthors{Berg\'e, Amara \& R\'efr\'egier}

\begin{document}
\title{Optimal capture of non-Gaussianity in weak lensing surveys : power spectrum, bispectrum and halo counts}
\author{Joel Berg\'e\altaffilmark{1,2}, Adam Amara\altaffilmark{3} and Alexandre R\'efr\'egier\altaffilmark{4}}
\altaffiltext{1}{Jet Propulsion Laboratory / California Institute of Technology, 4800 Oak Grove Drive, MS 169-327, Pasadena, CA 91109, USA ; e-mail: Joel.Berge@jpl.nasa.gov}
\altaffiltext{2}{NASA Postdoctoral Program (NPP) fellow}
\altaffiltext{3}{Department of Physics, ETH Zurich, Wolfgang-Pauli-Strasse16, CH-8093 Zurich, Switzerland}
\altaffiltext{4}{Laboratoire AIM, CEA/DSM - CNRS - Universit\'e Paris Diderot, DAPNIA/SAp, 91191 Gif-sur-Yvette, France}

\begin{abstract}
We compare the efficiency of weak lensing-selected galaxy clusters counts and of the weak lensing bispectrum at capturing non-Gaussian features in the dark matter distribution. We use the halo model to compute the weak lensing power spectrum, the bispectrum and the expected number of detected clusters, and derive constraints on cosmological parameters for a large, low systematic weak lensing survey, by focusing on the $\Omega_m$-$\sigma_8$ plane and on the dark energy equation of state. 
We separate the power spectrum into the resolved and the unresolved parts of the data, the resolved part being defined as detected clusters, and the unresolved part as the rest of the field. We consider four kinds of clusters counts, taking into account different amount of information : signal-to-noise ratio peak counts; counts as a function of clusters' mass; counts as a function of clusters' redshift; and counts as a function of clusters' mass and redshift. We show that when combined with the power spectrum, those four kinds of counts provide similar constraints, thus allowing one to perform the most direct counts, signal-to-noise peaks counts, and get percent level constraints on cosmological parameters.
We show that the weak lensing bispectrum gives constraints comparable to those given by the power spectrum and captures non-Gaussian features as well as clusters counts, its combination with the power spectrum giving errors on cosmological parameters that are similar to, if not marginally smaller than, those obtained when combining the power spectrum with cluster counts. We finally note that in order to reach its potential, the weak lensing bispectrum must be computed using all triangle configurations, as equilateral triangles alone do not provide useful information.
The appendices summarize the halo model, and the way the power spectrum and bispectrum are computed in this framework.
\end{abstract}

\keywords{cosmological parameters - large-scale structures - weak gravitational lensing}


\section{Introduction} \label{sect_intro}

Since its first detections in the early 2000's (\citealt{bacon00,vw00,wittman00}), weak gravitational lensing, the coherent distortion of distant galaxies by intervening dark matter, has become a premier tool to constrain the cosmological model (for reviews, see e.g. \citealt{mellier99,bartelmann01,refregier03,hoekstra08,munshi08}) and has been shown to be the most promising probe of dark energy (\citealt{detf,peacock06}). Surveys' optimization and systematics minimizations in both software and hardware have been investigated (\citealt{step1,step2,amara07,amara08,paulin08,amara09,great08}), favoring wide surveys, with well-controlled, stable Point Spread Function, with comprehensive photometric redshift follow-up. Those characteristics are shared by ambitious upcoming large area surveys, such as the Large Synoptic Survey Telescope (LSST)\footnote{http://www.lsst.org}, the Panoramic Survey Telescope \& Rapid Response System (Pan-STARRS)\footnote{http://pan-starrs.ifa.hawaii.edu}, Euclid\footnote{http://sci.esa.int/science-e/www/area/index.cfm?fareaid=102, http://www.euclid-imaging.net} and the Joint Dark Energy Mission (JDEM)\footnote{http://jdem.gsfc.nasa.gov}. They will provide us with a large amount of high quality imaging, well fitted to measure dark energy with weak lensing. 
Anticipating a significant improvement in removing systematics, we are left with the question of how to best extract the cosmological information out of the data. 
For instance, one must decide how to optimally capture non-Gaussianities and break the degeneracies between cosmological parameters as constrained by the extensively studied and measured weak lensing power spectrum. In this paper, we ignore primordial non-Gaussianities, and consider non-Gaussianities due to the growth of structures only.

The power spectrum, the Fourier transform of the 2-point correlation function of the shear field, has been the most used measurement so far, both from ground (e.g. \citealt{massey05,vw05,jarvis06,hoekstra06,semboloni06,benjamin07,fu08}) and from space (e.g. \citealt{rhodes04,heymans05,schrabback07,cosmos3d}). The introduction of tomography, the three-dimensional information of the shear field, that captures structure evolution, has tightened constraints on cosmological parameters, in particular the matter density $\Omega_m$ and the amplitude of density fluctuations $\sigma_8$ (\citealt{cosmos3d}). Despite its success, the power spectrum leaves us with well known degeneracies between parameters, that we must break by combining it with other measurements and/or probes. The power spectrum capturing only the Gaussian features of the field, it is natural to introduce measures of non-Gaussianities, and combine them with the power spectrum, to break degeneracies and tighten constraints.

Higher-order correlations measure the matter density field's non-Gaussian features. The lowest one, the weak lensing 3-point correlation function (3PCF), and its Fourier transform, the bispectrum, have been given a lot of attention in the past few years (\citealt{schneider03,tj03a,tj03b,tj03c,zaldarriaga03,tj04,schneider05,benabed06,semboloni08,joachimi09,vafaei09}). However, a clean measurement of the 3PCF is difficult, thus few papers have reported observational measurements so far. \cite{bernardeau03} and \cite{jarvis04} have measured the skewness, but a full measurement of the shear 3PCFs, or of the convergence bispectrum, is still to be done. Several efforts are underway, including an algorithm to measure the convergence bispectrum directly in Fourier space (\citealt{fastlens}).

Large-scale structures such as clusters of galaxies, the result of the non-linear evolution of density fluctuations, are the non-Gaussian features that we want to take into account. Assessing their abundance as a function of various parameters, such as redshift, has been known as a powerful probe and used as this for several years (e.g. \citealt{oukbir92,eke98,wang98,holder01,refregier02,battye03,pierpaoli03,wang04,horellou_berge05,marian06,gladders07,mantz08,sahlen08}). In particular, tight constraints can be obtained on $\Omega_m$ and $\sigma_8$. Because the weak lensing clusters selection function is rapidly evolving with redshift, counting weak-lensing-selected clusters is less sensitive than using catalogs of X-ray, optical, or Sunyaev-Zel'dovich (SZ) clusters. Despite this fact, constraining cosmology is possible with weak-lensing clusters only (\citealt{wk03,marian06,dietrich09,kratochvil09,marian09,maturi09,wang09}), as shown on real data by \cite{berge08}. Although such counts give weaker constraints than the weak lensing power spectrum, combining them with the power spectrum efficiently breaks degeneracies (\citealt{pires_model}).

In this paper, we show how combining the weak lensing power spectrum with the weak lensing bispectrum or counts of weak-lensing-selected clusters of galaxies will allow one to tighten constraints on cosmological parameters with upcoming weak lensing surveys. In particular, we show how we will be able to measure the equation of state $w=P/\rho$, where $P$ is the pressure, and  $\rho$ the density of the dark energy, down to the percent level. While other authors studied the combination of the power spectrum with cluster counts (e.g. \citealt{takada07}) and with the bispectrum (e.g. \citealt{tj04}), we present, for the first time, a detailed and consistent comparison of both combinations at once. 

Constraints on cosmological parameters have been studied by several authors (e.g. \citealt{cooray_hu01,tj04,takada07}). Most constraints' predictions have so far been made using a fitting function for the non-linear power spectrum, such as \cite{peacock96,ma00} or \cite{smith03}. However, fitting functions tuned for a $\Lambda$CDM universe must be used with caution when one is varying $w$ (\citealt{joudaki09}).
We use the halo model to compute the weak lensing statistics. Despite its simplicity, it is well suited to cosmological parameters prospects, since it allows us to vary the dark energy without extrapolating fitting functions out of $w=-1$. Appendix \ref{sect_hm} summarizes the halo model description that we use in this paper.
The halo model allows us to define the power spectrum as the separated contributions of the resolved and unresolved parts of the weak lensing field. Such a separation, since it gets rid of highly non-Gaussian features, can help improve constraints on cosmological parameters. We consider four different types of cluster counts, with more or less intrinsic cosmological information : counts as a function of shear signal-to-noise ratio (S/N) only, counts as a function of mass, counts as a function of redshift, and counts as a function of mass and redshit. We show that when combined with the power spectrum, counting clusters just as a function of their S/N gives constraints similar to those obtained when combining the power spectrum with counts of clusters taking the full mass and redshift information. We then show how the weak lensing bispectrum captures non-Gaussian features and breaks the power spectrum degeneracies as well as clusters counts. We assume a Euclid-like survey, 20,000 deg$^2$ wide, with $n_g=40 ~\mbox{galaxies arcmin}^{-2}$ with median redshift $z_m=1$, without any external (e.g. from CMB) priors, so as to show what can be done with a weak lensing survey alone.

Section \ref{sect_method} summarizes the methods we employ to constrain cosmological parameters with different weak lensing statistics. Section \ref{sect_results} presents our results. We conclude in section \ref{sect_ccl}.  Our halo model code will be made part of the public icosmo package\footnote{http://www.icosmo.org} (\citealt{icosmo}).


\section{Method} \label{sect_method}

\subsection{Weak lensing selection function} \label{sect_selfct}

The weak lensing selection function for clusters of galaxies has already been extensively studied, by using different kinds of smoothing functions or matched-filters (e.g. \citealt{wk02,hamana04,hennawi05,maturi05,marian06}). 
We derive a simplified, ideal selection function, based on a matched-filter approach for the signal-to-noise ratio created by a halo. We neglect projection effects and intrinsic alignments, which have been investigated e.g. by \cite{maturi05}, \cite{marian06} and \cite{fan07}. That selection function has been introduced, without the mathematical justification that follows, and tested on real data in \cite{berge08}.

We consider a spherically symmetric weak lens, decoupled from its surrounding and alone along the line of sight, characterized by its convergence $\kappa_{\rm obs}(\theta)$, where $\theta$ is the distance from its center (the convergence can be replaced by the shear without any loss of generality). The number density of background sources is $n_g$. We assume that the noise $n(\theta)$, originating from Poisson and intrinsic shape noise only, averages to 0 in circular shells and that its variance in the circular shell $i$ is given by $<n_i^2>=\sigma_\gamma^2/N_i$, where $\sigma_\gamma$ is the r.m.s error on the shape measurement and $N_i$ is the number of lensed galaxies in shell $i$. We fit a theoretical model $\kappa(\theta)$ to the observable $\kappa_{\rm obs}(\theta)=\kappa(\theta)+n(\theta)$.

The signal-to-noise ratio created by the lens is defined as the ratio of an estimator of $\kappa(\theta)$ to its associated error,  $\nu = \left< \hat{K}_w \right> / \sigma \left( \left< \hat{K}_w \right> \right)$. The estimator is defined by $\left< \hat{K}_w \right> = \int \dd^2 \theta \kappa_{\rm obs}(\theta) w(\theta)$, where $w(\theta)$ is a weight function to be adjusted so as to optimize the S/N. This is the case when $w=\kappa$, giving the optimized S/N as

\begin{figure}[t]
\centering
\includegraphics[width=8cm,angle=0]{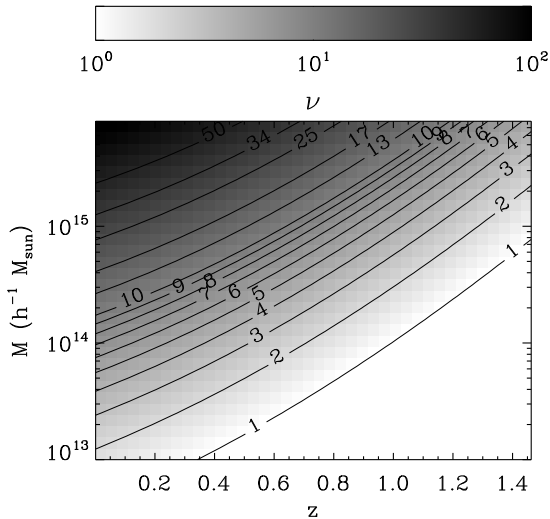} 
\caption{Weak lensing selection function for clusters of galaxies, in the redshift - mass plane. We assume a survey with $n_g=40$ galaxies per arcmin$^2$, distributed along equation \ref{nz}. Contours denote the S/N of galaxy clusters.} \label{fig_mlim}
\end{figure}

\begin{equation} \label{eq_nu1}
\nu = \frac{\sqrt{n_g}}{\sigma_\gamma} \sqrt{\int \dd^2 \theta \kappa^2(\theta)},
\end{equation}
which is consistent with previously published expressions.

We will consider NFW halos (\citealt{nfw96}) only, the S/N of which is expressed as (appendix \ref{app_nfw})
\begin{equation} \label{eq_snnfw}
\nu = 2 \sqrt{2 \pi} \left< Z \right> \frac{\sqrt{n_g}}{\sigma_\gamma} \frac{\rho_s r_s^2}{D_{\rm d} \Sigma_{\rm crit,\infty}} \sqrt{G(c)}
\end{equation}
where $G(c) = \int_0^c {\rm d}x ~x g(x)^2$ is a function of the halo's concentration $c$ only, well fitted by
\begin{equation}
G(c) \approx \frac{0.131}{c^2} - \frac{0.375}{c} +0.388 - 5 \times 10^{-4} c - 2.8 \times 10^{-7} c^2
\end{equation}
for $1 \leqslant c \leqslant 200$, and the function $g(x)$ is defined by Eq. (\ref{eq_g}). The quantity $\left<Z\right>$ in Eq. (\ref{eq_snnfw}) describes the effect of the distribution of galaxy sources. In the above equations, $D_{\rm d}$ is the angular-diameter distance to the lens, $\rho_s$ and $r_s$ parametrize the NFW halo (see Appendix \ref{sect_hm}), and $\Sigma_{\rm crit, \infty}$ is defined in Appendix \ref{app_nfw}.

We parametrize the redshift distribution of source galaxies by (\citealt{smail95})
\begin{equation} \label{nz}
n(z) = z^{\alpha} \exp \left[ -\left( \frac{z }{z_0} \right) ^{\beta} \right]
\end{equation}
with $\alpha=2$, $\beta=1.5$, and $z_0 \approx z_{\rm m}/1.412$, where $z_{\rm m}$ is the median redshift of the survey.

Figure \ref{fig_mlim} shows our fiducial survey's selection function, in the mass-redshift plane. Contours originate from the S/N of halos in that plane.

\subsection{Weak lensing statistics} \label{sect_stats}
\subsubsection{Power spectrum} \label{sect_ps}

The halo model allows us to decompose the full three-dimensional matter power spectrum along 
\begin{equation} \label{eq_gng}
P_\delta(k)=P_\delta^{\rm res}(k)+P_\delta^{\rm unr}(k),
\end{equation}
where $P_\delta^{\rm unr}(k)$ is computed on the {\it unresolved} part of the halos distribution, defined as the ensemble of halos with S/N smaller than a certain threshold $\nu_{\rm th}$; $P_\delta^{\rm res}$ is computed on the {\it resolved} part of the halos distribution, the ensemble of halos with $\nu > \nu_{\rm th}$. Those names come naturally from the separation made on a mass map between peaks (defined as resolved structures with $\nu>\nu_{\rm th}$) and the rest, unresolved part of the map. The resolved part can be seen as the most non-Gaussian features of the field.

\begin{figure}[t]
\centering
\includegraphics[width=8cm,angle=0]{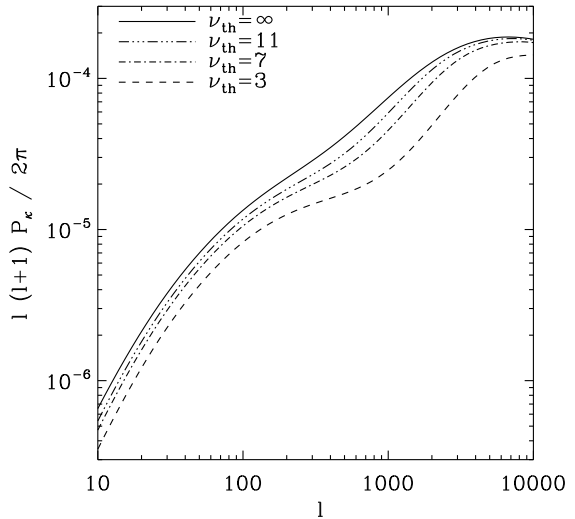} 
\caption{Weak lensing convergence power spectrum, when one single redshift bin is used to measure it. The black curve shows the power spectrum when the contribution of all halos is taken into account. Other curves make use of halos whose signal-to-noise is less than a given threshold (the unresolved part) : $\nu_{\rm th}=3$ (dashed), $\nu_{\rm th}=7$ (dash-dot) and $\nu_{\rm th}=11$ (dash-dot-dot-dot).} \label{fig_ps}
\end{figure}

The full weak lensing power spectrum $P_{\kappa}(\ell)$ follows the same decomposition. Figure \ref{fig_ps} shows the full power spectrum together with power spectra computed on unresolved parts defined by the thresholds $\nu_{\rm th}=3$, $\nu_{\rm th}=7$ and $\nu_{\rm th}=11$. The effect of removing significant clusters is mostly apparent on intermediate scales. It is less pronounced than when cutting the power spectrum with respect to halo masses, as shown by \cite{takada07}, because of the weak lensing selection function, that mixes masses when integrating the matter power spectrum on redshift.

The power spectrum is a measure of the Gaussian field only, and one must rely on direct measures of non-Gaussianity to complement it, like the bispectrum or cluster counts.

\subsubsection{Bispectrum} \label{sect_bs}

Three point statistics are the lowest statistics to capture non-Gaussian features in a statistical field. 
The dark matter and the weak lensing bispectra have been the object of numerous studies (e.g. \citealt{tj04}), most commonly based on the perturbation theory on large scales and on the hyper-extended perturbation theory (HEPT) on smaller scales (e.g. \citealt{bernardeau02} for a review). 
Here, we compare and combine the constraints from the weak lensing power spectrum and bispectrum evaluated in the same halo model. This fills a gap in the literature, where such constraints come from the HEPT approximation only.

The halo model allows us to separate the contribution from the unresolved and the resolved parts of the field when computing the bispectrum, like we did for the power spectrum. However, discarding the unresolved part from the bispectrum removes too much useful signal, and thus suppresses much power in the bispectrum. Therefore, we will always evaluate the bispectrum on the entire (resolved and unresolved) data field.

\subsubsection{Weak-lensing-selected halo counts} \label{sect_cc}

Clusters of galaxies, tracing dark matter halos, have emerged from the primordial Gaussian field through non-linear gravitational clustering, and are those non-Gaussianities that one can detect with higher order statistics. Or one can simply estimate their abundance to take non-Gaussianities into account. 
In this paper, we concentrate on how to break the degeneracies from the power spectrum by combining it with non-Gaussian features. As a result, we will not focus on the constraints that halo counts alone can bring, but on how they improve the constraints from the power spectrum.

Weak-lensing-selected clusters are detected on a mass map derived from the data used to measure the power spectrum and the bispectrum. Hence, they come at no extra cost, contrary to X-ray or optically-selected clusters. Defined as peaks with S/N greater than a certain threshold $\nu_{\rm th}$, they are the resolved part introduced above.

Once halos are selected, one can add information to their single S/N distribution, e.g. by introducing their redshift and mass, and thus measure their abundance as a function of those quantities.
Hereafter, we investigate the constraints on cosmological parameters provided by four kinds of halo counts. With increasing information, we look at (1) counts as a function of halo's S/N only ; (2) counts as a function of halo's mass ; (3) counts as a function of halo's redshift ; and (4) counts as a function of halo's redshift and mass.

In the remainder of this paper, we assume that the weak lensing selection function is perfectly known. We neglect the nuisance parameters commonly used to account for the imperfect knowledge of the relation between an observable and a cluster's mass (e.g. \citealt{hu03,lima04,majumdar04}).


\subsection{Constraints on cosmological parameters} \label{sect_fisher}

Fisher matrices allow one to characterize how a set of observables $\mathbf{x}$ is able to constrain a set of parameters $\mathbf{p}$ around a fiducial model. The associated Fisher matrix is defined by (e.g. \citealt{hutegmark99})
\begin{equation}
F_{\alpha \beta} = - \left< \frac{\partial^2 \ln L}{\partial p_\alpha \partial p_\beta} \right>.
\end{equation}
where $L(\mathbf{x},\mathbf{p})$ is the associated likelihood.

Given a fiducial model, the inverse of the Fisher matrix $\mathbf{F}^{-1}$ estimated in its neighborhood provides a lower limit to the parameters' covariance matrix. It thus quantifies the best statistical error that can be reached on the parameters, $\sigma(p_\alpha) \geqslant \sqrt{(\mathbf{F}^{-1})_{\alpha \alpha}}$, where $\sigma(p_\alpha)$ is the 1$\sigma$ error on parameter $p_\alpha$ marginalized over other parameters $p_\beta$. 

The Fisher matrices for the weak lensing power spectrum, bispectrum and cluster counts can be found e.g. in \cite{cooray01,lima04,hu_jain04,tj04} and \cite{takada07}.

We aim to determine how to best capture non-Gaussianities in order to break the degeneracies from the power spectrum alone. We thus need to combine the power spectrum with the bispectrum or the halo counts. 
To combine two observables ${\rm D_1}$ and ${\rm D_2}$, we create the data vector ${\rm D}=\{{\rm D_1},{\rm D_2}\}$.
The associated Fisher matrix is
$F_{\alpha \beta}^{({\rm (1)+(2)})} = {\rm D}_{,\alpha}^T \left[{\rm C^{((1)+(2))}}\right]^{-1} {\rm D}_{,\beta}$.

The rigorous estimation of the combined observables' covariance matrix,
\begin{equation}
{\rm C^{((1)+(2))}} =
\left(
\begin{array}{cc}
{\rm C^{(1)}} & {\rm C^{(1),(2)}} \\
{\rm C^{(1),(2)}} & {\rm C^{(2)}}
\end{array}
\right),
\end{equation}
requires the knowledge of the cross-covariance ${\rm C^{(1),(2)}}$. 

The weak lensing power spectrum is dominated by the most massive halos' contribution, those which are the most likely to be detected and taken into account in cluster counts. Hence, the power spectrum and halo counts are not independent observables : their cross-covariance is non-zero, and must be accounted for. 
Nonetheless, \cite{takada07} showed that neglecting it changes the errors by only a few percent.

Here, we take a different approach, by separating the contribution of different halos to the power spectrum (Eq. \ref{eq_gng}). We combine the counts of clusters with S/N greater than the threshold $\nu_{\rm th}$ (the resolved part) with the power spectrum created by all halos with S/N smaller than the same threshold $\nu_{\rm th}$ (the unresolved part). If we neglect the clustering between halos, the power spectrum in the unresolved part and the resolved clusters are uncorrelated : the covariance between cluster counts and the power spectrum vanishes, and we can simply add the Fisher matrices,
$F_{\alpha \beta}^{({\rm c+ps^{\rm unr}})} = F_{\alpha \beta}^{\rm c} + F_{\alpha \beta}^{\rm ps^{\rm unr}},$
where the superscript $^{\rm ps^{\rm unr}}$ stands for the power spectrum evaluated on the unresolved part.

As shown by \cite{tj04}, the cross-covariance between the power spectrum and the bispectrum has no Gaussian feature, but arises from the five-point correlation function of the shear field. In such a case, it is safe to approximate the Fisher matrix of the combination between the power spectrum and the bispectrum by the sum of the individual Fisher matrices,
$F^{\rm (ps+bisp)}_{\alpha \beta} \approx F^{\rm ps}_{\alpha \beta}+ F^{\rm bisp}_{\alpha \beta}.$


\section{Results} \label{sect_results}

\begin{figure*}[t]
\centering
\includegraphics[width=15cm,angle=0]{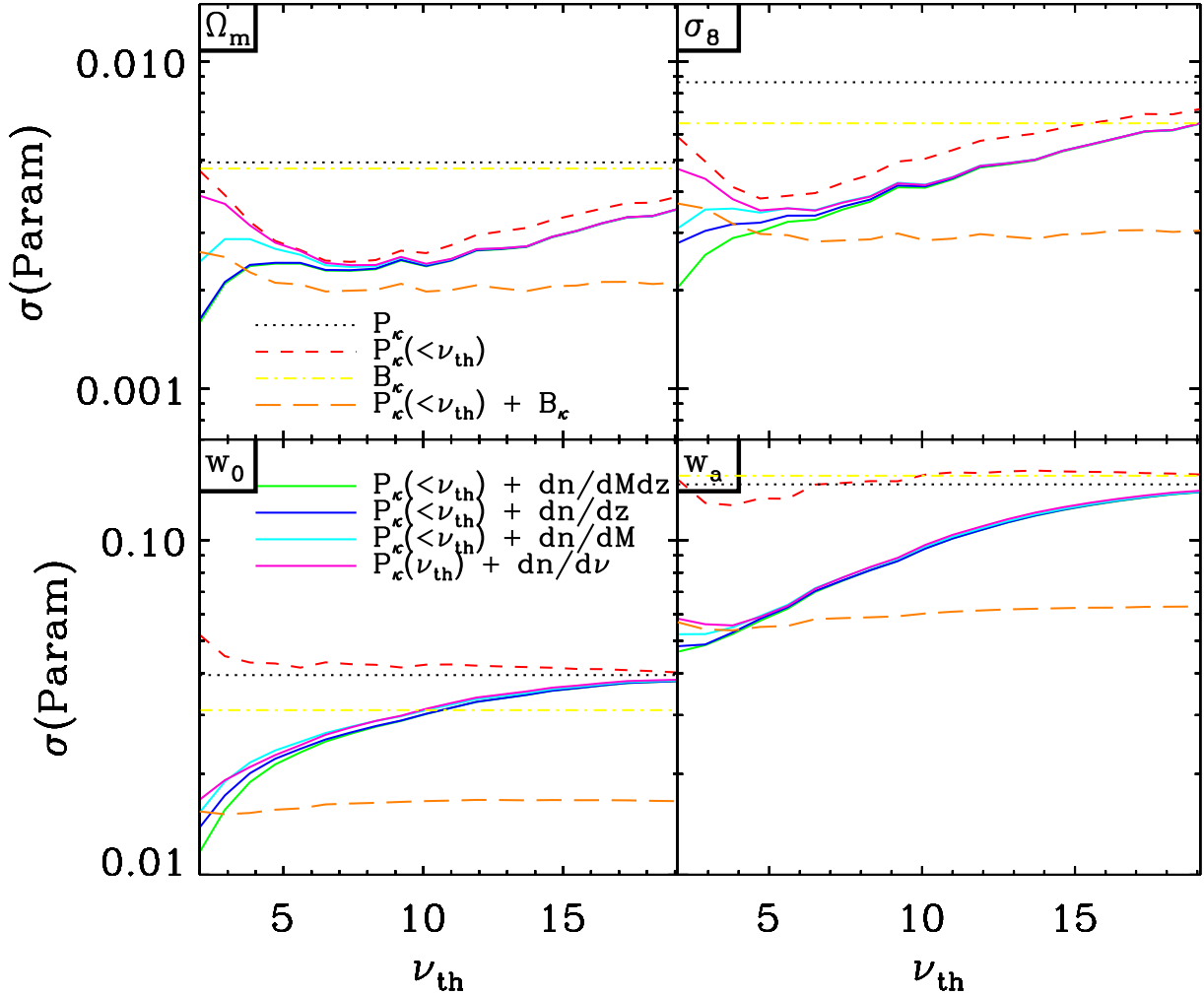}
\caption{Marginalized errors on cosmological parameters, as a function of the threshold between resolved and unresolved parts. Top-left: $\Omega_m$. Top-right : $\sigma_8$. Bottom-left : $w_0$. Bottom-right : $w_a$. Different colors and line styles label different measurements : power spectrum on all the data (dotted black), power spectrum on the unresolved part (dashed red), bispectrum (dash-dotted yellow), and combinations of the power spectrum on the unresolved part with clusters counts as a function of redshift and mass (solid green), redshift only (solid blue), mass only (solid cyan), S/N only (solid purple) and with the bispectrum (long-dashed orange). Note that the $y$-scale is not the same for top and bottom panels.} \label{fig_errors}
\end{figure*}

In this section, we compute and compare the errors on cosmological parameters, that can be reached by measuring the power spectrum alone, and when combining it with either counts or the bispectrum. We assume a fiducial cosmology described by the parameters $(\Omega_m,~\Omega_\Lambda,~\Omega_b,~\sigma_8,~h,~n,~w_0,~w_a)=(0.3,0.7,0.04,0.9,0.7,1,-1,0)$. The Universe's curvature is a free parameter, set by the combination of $\Omega_m$ and $\Omega_\Lambda$. The evolution of the dark energy equation of state is parametrized by $w(a)=w_0+(1-a)w_a$, $a$ being the scale factor.
We assume a Euclid-like survey, 20,000 deg$^2$ wide, with $n_g=40 ~\mbox{galaxies arcmin}^{-2}$ with median redshift $z_m=1$. We assume that the intrinsic shape r.m.s is $\sigma_{\rm int}=0.3$.

To count halos, we use 20 linear redshift bins, spanning the entire interval accessible to our fiducial survey ($0<z<5$) - the highest bins being empty - and 15 logarithmic mass bins ($10^{10} {\rm h}^{-1}{\rm M}_\odot<M<10^{16} {\rm h}^{-1}{\rm M}_\odot$ - where the lower bound is small enough so that all clusters detectable by their weak lensing signal are accounted for) to compute halo counts as a function of redshift and/or mass, and 15 linear S/N bins for S/N peak counts ($2<S/N<20$). We estimate the power spectrum in 10 redshift bins ($0<z<5$). For numerical reasons, we cannot use more than 3 redshift bins to compute the tomography bispectrum. This is not problematic, since \cite{tj04} showed that the tomography bispectrum S/N quickly converges to its maximum value, and almost reaches it for three redshift bins.

Before reporting our results, we would like to refer to the work of \cite{vallisneri08} in which he shows the limitations of the Fisher matrix formalism. In particular, one must pay particular attention to ill-conditioned Fisher matrices, the inversion of which is likely to be wrong. Here, we check the condition number of all of our Fisher matrices, and consider as good only those with a small enough condition number so that their inversion can be trusted. We find that the Fisher matrices for counts as a function of S/N, of mass and of redshift, are all ill-conditionned. Therefore, such halo counts cannot give reliable Fisher constraints by themselves. This is true for our particular set of parameters though, and could not be true on other parameter spaces. Looking for such spaces is beyond the scope of this paper, and we will give constraints from halo counts only when combined with the power spectrum, for which the condition number is low enough to be safe.

\begin{table*}[t]
\caption{Marginalized absolute errors on cosmological parameters.} \label{table_constraints}
\begin{center}
\begin{tabular}{ccccccccccc}
\tableline\tableline
 & $P_\kappa$  & $P_\kappa^{\rm unr}$ & $B_\kappa$ & $n(M,z)$ & $P^{\rm unr}_\kappa+n(M,z)$ & $P^{\rm unr}_\kappa+n(z)$ & $P^{\rm unr}_\kappa+n(M)$ & $P^{\rm unr}_\kappa+n(\nu)$ & $P^{\rm unr}_\kappa+B_\kappa$ & $P_\kappa+B_\kappa$ \\
\hline
$\Omega_m$ (0.3) & 0.0049 & 0.0027 & 0.0047 & 0.070 & 0.0024 & 0.0024 & 0.0026 & 0.0026 & 0.0021& 0.0023 \\
$\Omega_\Lambda$ (0.7) & 0.024 & 0.012 & 0.019 & 0.254 & 0.0099 & 0.010 & 0.011& 0.011 & 0.0087 & 0.0094 \\
$\Omega_b$ (0.04) &  0.015 & 0.015 & 0.016 & 0.656 & 0.013 & 0.014 & 0.014 & 0.014 & 0.008 & 0.008\\
$\sigma_8$ (0.9) &  0.0086 & 0.0039 & 0.0065 & 0.112 & 0.0032 & 0.0034 & 0.0036 & 0.0036 & 0.0029 & 0.0033\\
$h$ (0.7) &  0.091 & 0.086 & 0.122 & 3.59 & 0.077 & 0.079 & 0.079 & 0.085 & 0.048 & 0.051\\
$w_0$ (-1) &  0.040 & 0.041 & 0.031 & 0.283 & 0.023 & 0.024 & 0.025 & 0.025 & 0.016 & 0.016 \\
$w_a$ (0) &  0.147 & 0.133 & 0.156 & 2.179 & 0.062 & 0.063 & 0.064 & 0.064 & 0.055 & 0.063 \\
$w_p$ &  0.022 & 0.015 & 0.013 & 0.252 & 0.014 & 0.014 & 0.015 & 0.015 & 0.0049 & 0.0045 \\
$n$ (1) &  0.020 & 0.021 & 0.038 & 1.00 & 0.018 & 0.018 & 0.019 & 0.020 & 0.011& 0.012 \\
\tableline\tableline

\multicolumn{10}{l}{The S/N threshold between the resolved and unresolved parts is set to $\nu_{\rm th}=6$.} \\
\multicolumn{10}{l}{The central value for the parameter in our fiducial model is given into parenthesis in the first column.} \\
\multicolumn{10}{l}{We ignore combining the power spectrum with the bispectrum and cluster counts at a time (see main text).}
\end{tabular}
\end{center}
\end{table*}

The goal of this paper is to compare how cluster counts and the bispectrum capture non-Gaussianities and break degeneracies left over by the power spectrum. Consequently, we ignore combining cluster counts with the bispectrum, as well as combining the three observables (power spectrum, bispectrum and cluster counts) at a time.


We evaluate Fisher matrices for power spectra estimated both on all the data and on the unresolved parts defined by different $\nu_{\rm th}$, for the corresponding halo counts (halos with $\nu > \nu_{\rm th}$), and for the bispectrum, and the combinations introduced in section \ref{sect_fisher}. 
Figure \ref{fig_errors} shows the expected errors that they provide, marginalized over all eight parameters, as a function of the S/N threshold $\nu_{\rm th}$. The power spectrum $P_\kappa(<\nu_{\rm th})$ is estimated on the unresolved part (dashed red), then combined with clusters counts of clusters (made as a function of mass and redshift - solid green-, redshift only -solid blue-, mass only -solid cyan- and S/N only -solid purple). The dotted black flat line shows the errors brought by the measurement of the power spectrum on the entire data set. Discarding the contribution of the most massive halos (i.e. the most non-Gaussian features) from the power spectrum improves the errors on $\Omega_m$ and $\sigma_8$, the errors being smallest for an optimal threshold $\nu_{\rm th,opt} \approx 6$. Although the counter-intuitive increase in the performance that we observe when discarding the resolved part of the data is not so surprising, since the impact of non-Gaussianities on the power spectrum is lowered  (\citealt{shaw09} observe a similar trend for Sunyaev-Zel'dovich clusters), we should note here that it could be exacerbated if our analysis is close to the Fisher matrices formalism's limits. Moreover, there may be an optimal weighting scheme that would both allow for the signal in the resolved part and minimize its non-Gaussian errors. Smaller errors on the cosmological parameters would then be expected, since more information (with minimal noise) would be taken into account than that we use here with the unresolved part. Investigating these issues is beyond the scope of this paper, and we defer it to a later study. The dark energy equation of state parameters $w_0$ and $w_a$ are less affected by the separation of the power spectrum into the resolved and the unresolved parts.

Although cluster counts do not provide strong constraints by themselves (see below, Table \ref{table_constraints}), combining them with the power spectrum improves the errors it provides, in particular when low S/N clusters are taken into account. When the clusters considered are significant enough, all four kinds of counts perform as well at capturing non-Gaussianities : the power spectrum provides most of the Fisher information, and clusters help by breaking parameters degeneracies, independently of the way they are considered. Figure \ref{fig_errors} shows that this is true for $\nu_{\rm th} \gtrapprox 6$, a safe threshold to discard false detections (\citealt{pace07}). Therefore, that makes S/N peaks counts a direct and efficient probe to combine with the power spectrum, as shown e.g. by \cite{pires_model}.

The dash-dotted yellow line on Fig. \ref{fig_errors} shows the errors given by the bispectrum, and the long-dashed orange line shows those given when combining it with the power spectrum. 
At low S/N thresholds, when combined with the power spectrum, the bispectrum and clusters counts give similar constraints on the four parameters $\Omega_m$, $\sigma_8$, $w_0$ and $w_a$. At higher thresholds the bispectrum gives better constraints than clusters counts on the dark energy parameters, yet errors remain comparable for $\Omega_m$ and $\sigma_8$.

Table \ref{table_constraints} lists the marginalized errors on all of our eight parameters, as well as those of the dark energy equation of state $w_p$ at the pivot point (section \ref{sect_de}), for the measurements and combinations considered, with $\nu_{\rm th}=6$. Halo counts (excepted counts as a function of redshift and mass) are plagued by ill-conditioned Fisher matrices, and we do not consider them alone. Nevertheless, since counts as a function of S/N, of mass alone and redshift alone contain less information than counts as a function of mass and redshift, their expected constraints are weaker than those listed in table \ref{table_constraints} for counts as a function of mass and redshift.
We also list in the table the errors provided when combining the bispectrum with the power spectrum estimated on all the data. Although this combination is not explicitly shown on figure \ref{fig_errors}, it can be drawn from the high S/N threshold-limit of the combination of the bispectrum with the power spectrum estimated on the unresolved part (long-dashed orange line). 
The table emphasizes the conclusions given by figure \ref{fig_errors}. The four combinations of clusters counts with the power spectrum give similar constraints, that are comparable to those provided by the combination of the bispectrum and the power spectrum. Furthermore, although the bispectrum and the power spectrum give comparable errors, combining them breaks degeneracies and lowers errors by a factor $\approx 2$.

\cite{pires_model} looked at how several statistics (clusters counts, bispectrum, skewness, kurtosis) break the $\Omega_m$-$\sigma_8$ degeneracy, and reported the high efficiency of cluster counts. They found that the skewness could also discriminate against models, but that the bispectrum, when using only equilateral triangles, gave poor results.
We expand their work by taking into account all triangle configuration when computing the bispectrum, and add it to clusters counts as an efficient way to break degeneracies (see section \ref{sect_csand}). In particular, Fig. \ref{fig_errors} shows that this conclusion is true not only for the $\Omega_m$-$\sigma_8$ degeneracy, but also for the dark energy equation of state.

\subsection{Dark energy plane} \label{sect_de}

\begin{figure}[t]
\centering
\includegraphics[width=8cm,angle=0]{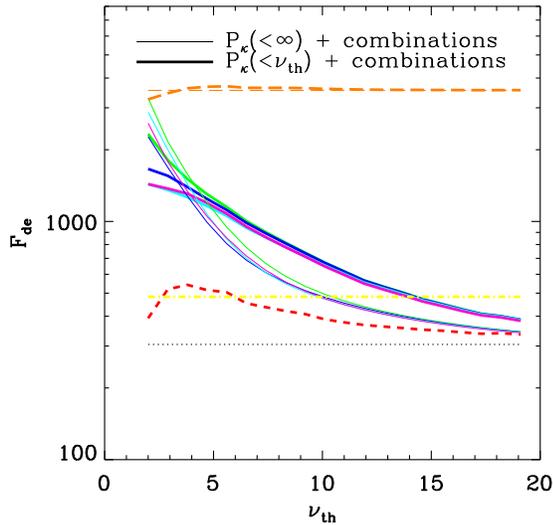}
\caption{Dark energy FoM, as defined by Eq. (\ref{eq_fomw}), as a function of the threshold between the resolved and the unresolved parts, for the same measurements as in Fig. \ref{fig_errors}, labeled with the same colors and line styles. Thick lines denote measurements made with the power spectrum estimated on the unresolved part, and thin lines take the entire information into account in the power spectrum estimation.} \label{fig_fom_detf}
\end{figure}

We now specialize the discussion to the dark energy figure of merit, as defined e.g. by \cite{albrecht09},
\begin{equation} \label{eq_fomw}
\mathcal{F}_{\rm de}=\frac{1}{\sigma(w_p)\sigma(w_a)}
\end{equation}
where $w_p$ is the dark energy equation of state at the pivot redshift, where the dark energy is best constrained. The errors on $w_p$ are listed in Table \ref{table_constraints}.

Figure \ref{fig_fom_detf} shows the dependence of this figure of merit as a function of the S/N threshold $\nu_{\rm th}$ between the resolved and the unresolved parts of the data. The color code and line styles are the same as those used for the previous figure. Thick lines show the FoMs when the power spectrum is estimated on the unresolved part, and thin lines show the FoMs when it is estimated on all the data (we neglect the covariance between the power spectrum and cluster counts).

Considering the power spectrum on the unresolved part only (dashed red) slightly improves the FoM. This holds true when combining clusters with $\nu \geqslant 5$. The bispectrum gives a higher figure of merit than cluster counts, and depends very weakly on $\nu_{\rm th}$. Only at low S/N do the figures of merit compare.

Clusters counts, when the selection function and the contamination are controlled well enough to allow one to consider a very low S/N threshold, and the bispectrum are comparable at extracting non-Gaussian aspects in a weak lensing survey and at constraining dark energy. On the other hand, if we use a safe threshold for clusters detection ($\nu_{\rm th} \approx 6$), the bispectrum becomes better than clusters counts to break degeneracies and to measure dark energy.

\subsection{Comparison with other works}

\begin{figure}[t]
\centering
\includegraphics[width=8cm,angle=0]{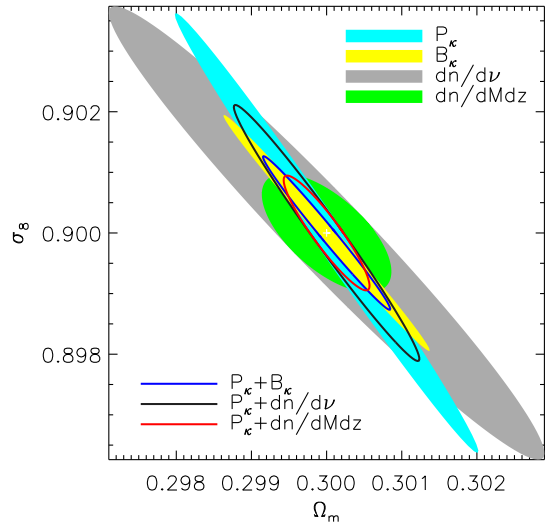}
\caption{Constraints on the $\Omega_m$-$\sigma_8$ plane, when all other parameters are kept constant, for a 20,000 deg$^2$ survey, and no tomography. Filled ellipses show constraints from the power spectrum (cyan), S/N peak counts (grey), cluster counts as a function of mass and redshift (green) and bispectrum when all triangle configuration are taken into account (yellow); the constraints obtained from the bispectrum when taking into account equilateral triangles only get out of the frame. Open ellipses are constraints from combining the power spectrum with the bispectrum (all triangle configurations - blue), with peak counts (black) and cluster counts (red).} \label{fig_dp}
\end{figure}

\subsubsection{Peak counting}

In two recent papers, \cite{dietrich09} and \cite{kratochvil09} show with numerical simulations how powerful weak lensing peaks counting is at constraining $\Omega_m$ and $\sigma_8$, and $w$, respectively. These appear to contradict our results, in which we saw that if indeed, counting peaks efficiently captures non-Gaussianity and breaks power spectrum degeneracies, it does it only when combined with the power spectrum, but gives poor constraints when used alone (Table \ref{table_constraints}). A concern expressed by the aforementioned authors is that they varied only their parameters of interest ($\Omega_m$ and $\sigma_8$ or $w$), and defer marginalizing on other parameters for later works. In our analysis, we indeed marginalized on eight parameters. To show that this considerably lowers the constraints, we make the same kind of analysis, by varying $\Omega_m$ and $\sigma_8$ only, as done by \cite{dietrich09}. For simplicity, we do not separate the power spectrum into resolved and unresolved parts, but consider the total power spectrum only, without tomography. Since our analytical model cannot fully reproduce their taking into account cosmologically significant false detections (peaks that do not correspond to a single, localized, virialized cluster, but that are instead due to mass alignments along the line of sight), we use counts as a function of S/N as a proxy of their measure. We use a detection threshold $\nu_{\rm th}=3$.
Figure \ref{fig_dp} shows how counting peaks (grey filled ellipse) compares to the power spectrum (cyan filled ellipse) in that case. The black open ellipse shows the combination of the two. As shown by \cite{dietrich09}, peak counts indeed is an efficient cosmological tool when varying $\Omega_m$ and $\sigma_8$ only. Marginalizing on other parameters hampers its ability to perform, as we showed above, unless combined with the power spectrum. 

However, since our model does not catch filaments and significant false detections, as \cite{dietrich09} and \cite{kratochvil09} do, our conclusion about cluster counts alone may be over-pessimistic. Therefore, it will be valuable to adapt their simulations to higher dimensional parameter estimations.

\subsubsection{Bispectrum : the need for all triangle configurations} \label{sect_csand}

Looking for the best statistics to discriminate models, \cite{pires_model} conclude that peak counting is the best way of breaking the $\Omega_m-\sigma_8$ degeneracy, with other third-order statistics (such the skewness) also being able to break degeneracies. In that work the bispectrum in particular was found to be weak at breaking degeneracies. This apparent discrepancy with our findings here, where we find that combining the bispectrum with the power spectrum is comparable to combining peak counting with the power spectrum,  comes from the fact that the bispectrum presented here uses all triangle configurations, while the bispectrum in \cite{pires_model} used only equilateral triangles.

To emphasize the importance of considering all triangle configurations when computing the bispectrum, we compute the constraints expected on $\Omega_m$ and $\sigma_8$, while keeping all other parameters fixed, as in \cite{pires_model}, from the power spectrum, the bispectrum with all triangles information or only equilateral triangles, and cluster counts. We use one redshift bin. Our analytical model does not allow us to compute \cite{pires_model}'s Wavelet Peak Counts (counts of clusters in different wavelet scales), so we use the counts as a function of mass and redshift as a proxy. 

Figure \ref{fig_dp} shows how constraints from these measurements compare. The cyan filled ellipse shows constraints from the power spectrum, the green one, constraints from cluster counts, and the red open ellipse their combination. As shown by \cite{pires_model}, cluster counts are very efficient at grabbing non-Gaussianities and breaking the power spectrum degeneracies in that particular case.
The yellow filled ellipse shows constraints from the bispectrum, when the information from all triangle configurations is taken into account, and the blue open ellipse shows constraints from combining this bispectrum with the power spectrum. As mentioned above, the bispectrum (with information from all triangle configurations) provides tight constraints, and is as efficient as cluster counts at capturing non-Gaussian features. However, we find that when taking into account only equilateral triangles when computing constraints from the bispectrum, the errors on the parameters reach 100\%, making such a measurement extremely low efficiency. The frame of Figure \ref{fig_dp} is much too small to show the associated ellipse. 
Consequently, we concur with \cite{pires_model} when claiming that the bispectrum measured with equilateral triangles only does not allow one to break the power spectrum degeneracy between $\Omega_m$ and $\sigma_8$.

Coming back to our 8-parameters analysis, we notice the same behavior as that mentioned in the above paragraph. To capture non-Gaussian features and efficiently break degeneracies left over by the power spectrum, one has to take into account all triangle configurations when measuring the weak lensing bispectrum. Equilateral triangles only do not provide any cosmological constraints.


\section{Conclusion} \label{sect_ccl}

We investigated how to best capture non-Gaussianities in weak lensing surveys so as to break degeneracies left over by the power spectrum in weak lensing analysis, by means of combining it with galaxy clusters counts and with the weak lensing bispectrum. To this end, we computed, in the halo model framework, the power spectrum, the bispectrum and four kinds of halo counts (as a function of S/N, mass, redshift, and mass and redshift), in a Euclid-like survey. We have computed the errors on cosmological parameters brought by the combination of the power spectrum with the bispectrum, and of the power spectrum with clusters counts. We have emphasized our discussion on the varying dark energy equation of state, the mass density parameter $\Omega_m$ and the amplitude of density fluctuations $\sigma_8$, marginalized on a 8-parameter cosmological model. We did not take any prior from outside weak lensing (e.g. CMB priors), so as to show how a weak lensing survey alone can allow us to capture non-Gaussianities, and what constraints we can expect from it on cosmological parameters.

When computing the power spectrum, we have separated the contribution of the resolved and the unresolved parts of the data, delimited by a S/N threshold $\nu_{\rm th}$. Doing so, we noted an improvement in errors on cosmological parameters.

Although peak counts alone are powerful at constraining  $\Omega_m$ and $\sigma_8$ when other parameters are kept constant (as shown by \citealt{dietrich09}), we have seen that marginalizing on a 8-parameter space make them provide only weak constraints, unless combined with the power spectrum.
We have indeed shown that when combined with the power spectrum, cluster counts provide percent level marginalized errors on cosmological parameters.
Furthermore, provided that the threshold is high enough ($\nu_{\rm th} \approx 6$), all four kinds of cluster counts we consider give similar constraints. Hence, combining the power spectrum with S/N peak counts gives constraints nearly as tight as combining the power spectrum with clusters counted as a function of redshift and mass. Galaxy clusters provide the same non-Gaussian information, independently of the way they are taken into account, as long as they are combined with the power spectrum : different cluster counts break degeneracies in a similar way. This will allow us to put stringent constraints on cosmological parameters from weak lensing alone, by combining S/N peaks with the power spectrum, without any extra cost and requirement of follow-up for clusters' redshift and mass measurement.

We have shown that the weak lensing bispectrum efficiently captures non-Gaussianities, and is marginally more efficient than cluster counts at breaking degeneracies left over by the power spectrum. Furthermore, the bispectrum alone provides constraints comparable to those given by the power spectrum.
We have expanded \cite{pires_model}'s conclusions by adding the bispectrum to clusters counts as efficient tools to capture non-Gaussianities.
We also noted that equilateral triangles alone do not bring significant information. To provide percent level constraints, all triangle configurations must be accounted for when measuring the weak lensing bispectrum.

All our results are based on a Euclid-like survey, and may not be directly applicable to current surveys such as the Canada-France-Hawaii Telescope Legacy Survey (CFHTLS) or the Cosmic Evolution Survey (COSMOS). In particular, on current surveys, the cosmic variance could play an important role in lowering the ability of the bispectrum to be more efficient than S/N peak counts. How those two measures compare to capture non-Gaussian features could then depend on the survey's characteristics. This question is beyond the scope of this paper, and will be addressed in a subsequent work.

We have assumed that all systematics were well accounted for so as not to bias our results. Although that can seem over-optimistic, the amount of efforts currently underway to correct for various systematics gives us good reasons to think that our assumption is likely to be met when a Euclid-like survey is undertaken.
Moreover, allowing for systematics requires a more sophisticated analysis than the Fisher matrix (\citealt{amara08}). We defer it to a future paper, looking at how systematics enter in the correlations between different statistics.

In light of our results, it appears that weak lensing surveys alone will be able to reach the percent accuracy on the dark energy equation of state parameters. Moreover, it provides us with ways to cross-check the parameters' measurements. For instance, after measuring the power spectrum, counting S/N peaks provides an easy and fast way to optimally capture non-Gaussianities and tighten constraints. Then, the bispectrum and its combination with the power spectrum should give consistent parameters' estimation and similar constraints. However, measuring the bispectrum on half the sky will constitute a real challenge. 
Along with already existing robust and fast algorithms (\citealt{jarvis04,zhang05}), new ones are being developed (e.g. Berg\'e et al in prep and Semboloni et al in prep) to measure either the 3PCF in real space or the bispectrum in Fourier space (as proposed by \citealt{fastlens}).
In the meantime, more and more precise numerical simulations will be a needed tool to test and tune those algorithms, and current surveys such as the CFHTLS and COSMOS are ideal benchmark to test codes in the real world.

\acknowledgements{We want to thank Sandrine Pires, Benjamin Joachimi, Xun Shi and Masahiro Takada for fruitful discussions and for their help in comparing bispectrum codes, as well as Jochen Weller. We also thank Jason Rhodes and Sedona Price for useful comments on this manuscript. 
JB is supported by the NASA Postdoctoral Program, administered by Oak Ridge Associated Universities through a contract with NASA.
This work was carried out at Jet Propulsion Laboratory, California Institute of Technology, under a contract with NASA.
We thank the anonymous referee for useful comments.}

\appendix

\section{A - Halo model} \label{sect_hm}

Presented as an alternative to fitting functions based on numerical simulations, the halo model for large-scale structures describes structures as spherical haloes  - see \cite{cooray_sheth02} for a review. In this framework, the large-scale structures statistics can be described by correlations within a same halo (the profile of which affects them) and between different halos (the clustering of halos is thus naturally taken into account). Although being simplistic, the model agrees fairly well with common numerical simulations.
Here, we summarize its key ingredients : the mass function, the profile of halos and the halo bias. We also give the expressions for the 3D dark matter power spectrum and bispectrum.

\subsection{Mass function}

The comoving number of virialized halos, whose mass ranges between $M$ and $M+\dd M$ and redshift ranges between $z$ and $z+\dd z$, is given by (\citealt{ps}) :
\begin{equation} \label{eq_nmz}
n(M,z) \dd M = \frac{\rho_0}{M} \frac{\dd \nu(M,z)}{\dd M} f(\nu) \dd M
\end{equation} 
where $\rho_0$ is the Universe's current background density and $\nu(M,z)=\delta_c(z)/\sigma(M)$.
\, $\delta_c(z)$ is the non-linear overdensity of a halo collapsing at redshift $z$. In a dark energy universe, $\delta_c(z)$ depends weakly on cosmology, especially on the dark energy's equation of state $w$. We use \cite{wk03}'s fitting function to compute $\delta_c(z;w)$. This fitting function is valid for constant $w$ only, therefore we compute it using an effective equation of state $w_{\rm eff}(a)=w_0+(1-a)w_a$. 
The r.m.s of the mass fluctuations in a sphere containing a mass $M$ at redshift $z$ is defined by
\begin{equation}
\sigma^2(M)=\frac{1}{(2\pi)^3} \int \dd^3k P^{\rm lin}(k) |W(kR)|^2
\end{equation}
where $P^{\rm lin}(k)$ is the three-dimension linear matter power spectrum, $R=(3M/4\pi\bar{\rho})^{1/3}$ is the radius of the considered sphere, and $\bar{\rho}$ is the mean background density. The window function $W(x)=(3/x^3)\left[ \sin(x)-x \cos(x) \right]$.

A couple of mass functions have been described in the literature (\citealt{ps,st99,jenkins01,tinker08}). We use that of \cite{jenkins01},
$f(\sigma) = a_j \exp (-|\ln \sigma^{-1}+b_j|^{c_j})$,
where $a_j=0.315$, $b_j=0.61$ and $c_j=3.8$.

\subsection{Profile}

The top-hat collapse model for structure formation describes the matter infall on the gravitational wells until virialization. At this stage, non-linear physics takes place, that further modifies the virialized halo profile, and that must be assessed through numerical simulation. \cite{nfw96} showed that the mass profile of such virialized halos can be described by a ``universal" profile,
\begin{equation}
\rho(r|M) = \frac{\rho_s}{(r/r_s)^\alpha (1+r/r_s)^\beta},
\end{equation}
where $r_s$ and $\rho_s$ correspond to a characteristic radius and density, respectively, and ($\alpha,~\beta$)=(1,2) for the usual NFW profile, that we use in this paper.

The mass of a NFW halo is then given by 
$M=\int_0^{r_{\rm vir}} \dd r 4 \pi r^2 \rho(r|M)$,
where $r_{\rm vir}$ is the halo's virial radius. In theory, a NFW is infinite, but we assume here that it is truncated at $r_{\rm vir}$. The virial and characteristic radius are linked by $r_{\rm vir}=c r_s$, where $c$ is the halo's concentration, that can be parametrized, following \cite{bullock01} and \cite{dolag04}, by
\begin{equation} \label{eq_concentration}
c(M,z)=\frac{c_0}{1+z} \left( \frac{M}{M_*(z=0)} \right) ^{-\beta(z)}.
\end{equation}
In this equation, $M_*(z=0)$ is the non-linear mass scale, defined as $\nu(M_*,z=0)=1$. 
In contrast to earlier works (\citealt{bullock01}), we add an extra redshift-dependence in the parameter $\beta$. Based on \cite{zhao03,zhao08,duffy08} and \cite{gao08}, it allows us to avoid the catastrophic drop in concentration observed on simulations for high redshift, massive halos, as parametrized by \cite{bullock01} with a constant $\beta$. We find that $c_0=8$ and $\beta(z)=0.13-0.05z$, besides agreeing with \cite{duffy08,gao08,zhao08}, and with \cite{bullock01} at low redshift, gives us the best agreement between the halo model and the \cite{smith03}'s power spectrum and bispectrum. One should note that the concentration has a significant impact on the 1-halo term of the halo model power spectrum and bispectrum, as shown by \cite{huffenberger03}, and thus a thorough description is needed.

\subsection{Bias}

The bias describes how halos cluster with respect to the matter distribution. It is well described at first order by fitting functions of the form (\citealt{mo_white,smt01}),
\begin{equation}
b(M,z)=1-\frac{q\nu-1}{\delta_c(z)}+\frac{2p/\delta_c(z)}{1+(q\nu)^p}
\end{equation}
with $q=0.707$ and $p=0.3$. We neglect the higher order terms of the bias.

\subsection{Matter power spectrum and bispectrum}

The halo model describes the correlation functions of the density field, in real-space, as the sum of the correlation between points belonging to a same halo and of the correlation between points in different halos. The same contributions appear in Fourier space, when defining the power spectrum and bispectrum, which are thus the sum of a 1-halo term and a 2-halo term (and 3-halo term for the bispectrum) (e.g. \citealt{cooray_hu01}),
$P_\delta(k) = P_\delta^{1h}(k)+P_\delta^{2h}(k).$

The 1-halo component is given by
\begin{equation}
P_\delta^{1h}(k) = \int \dd m n(m) \left( \frac{m}{\bar{\rho}} \right)^2 |u(k|m)|^2
\end{equation}
and the 2-halo term is given by
\begin{equation}
P_\delta^{2h}(k) = \int \dd m_1 n(m_1) \left( \frac{m_1}{\bar{\rho}} \right) u(k|m_1) 
	\int \dd m_2 n(m_2) \left( \frac{m_2}{\bar{\rho}} \right) u(k|m_2) P_{hh}(k|m_1m_2).
\end{equation}

The function 
\begin{equation} \label{eq_u} 
u(k|m) = \int_0^{r_{\rm vir}} \dd r 4\pi r^2 \frac{\sin kr}{kr} \frac{\rho(r|m)}{m}
\end{equation}
is the Fourier transform of the dark matter density profile.
The term $P_{hh}(k|m_1m_2)$ represents the power spectrum of halos of mass $m_1$ and $m_2$ and is approximated by $P_{hh}(k|m_1m_2) \approx \Pi_{i=1}^2 b_i(m_i) P_\delta^{\rm lin}(k)$, where $b_i(m_i)$ is halo $i$'s bias and $P_\delta^{\rm lin}(k)$ is the matter linear power spectrum. We use the \cite{eisenstein_hu}'s transfert function to evaluate $P_\delta^{\rm lin}(k)$.

The 3D matter bispectrum can be decomposed into terms coming from one, two and three halos :
\begin{equation}
B_\delta = B_\delta^{1h}+B_\delta^{2h}+B_\delta^{3h},
\end{equation}
with
\begin{equation}
B_\delta^{1h}(k_1,k_2,k_3)=I_3^0(k_1,k_2,k_3),
\end{equation}
\begin{equation}
B_\delta^{2h}(k_1,k_2,k_3) = I_2^1(k_1,k_2)I_1^1(k_3)P_\delta(k_3) + C.P.,
\end{equation}
and
\begin{equation}
B_\delta^{3h}(k_1,k_2,k_3)=[2J(k_1,k_2,k_3)I_1^1(k_3)+I_1^2(k_3)] \\
	\times I_1^1(k_1)I_1^1(k_2)P_\delta(k_1)P_\delta(k_2) + C.P.,
\end{equation}
where $C.P.$ denotes circular permutations.
The $J$ function is given by the second-order perturbation theory (\citealt{bernardeau02}) :
\begin{equation}
J(k_1,k_2,k_3) = 1 - \frac{2}{7} \Omega_m^{-2/63} + \left( \frac{k_3^2-k_1^2-k_2^2}{2k_1k_2} \right) ^2 \\
	\times \left[ \frac{k_1^2+k_2^2}{k_3^2-k_1^2-k_2^2} + \frac{2}{7} \Omega_m^{-2/63} \right].
\end{equation}

The function $I_\mu^\beta$ is defined through the Fourier transform of the halo profile $u(k|m)$ (Eq. \ref{eq_u}) :
\begin{equation}
I_\mu^\beta(k_1,\dots,k_\mu;~z) = \int dm \left( \frac{m}{\rho_b} \right)^\mu n(m,z)b_\beta(m) \\
\times u(k_1,m) \dots u(k_\mu,m)
\end{equation}
with $b_0=1$.

\section{B - Signal-to-noise ratio for a NFW halo} \label{app_nfw}

Here, we specialize the discussion of section \ref{sect_selfct} to an NFW halo. To evaluate equation (\ref{eq_nu1}) in this case, we first need to derive the convergence of such a halo.
The mass density, projected along the line-of-sight, of a NFW halo at redshift $z$, with concentration $c$, is given by :
\begin{equation}
\Sigma(x) = \int_{-\sqrt{c^2-x^2}}^{\sqrt{c^2-x^2}} {\rm d}z \rho(x,z) = 2 \rho_s r_s g(x)
\end{equation}
where $x=r/r_s$. The function $g$ depends only on the distance to the halo's center, and is given by (as long as $c \geqslant 1$) \citep{wright00}

\begin{equation} \label{eq_g}
g(x) = 
\left\{
\begin{array}{l}
-\frac{\sqrt{c^2-x^2}}{(1-x^2)(1+c)} + \frac{1}{(1-x^2)^{3/2}} {\rm arccosh} \frac{x^2+c}{x(1+c)}, \\ 
\hspace{47mm} (x < 1) \\
\frac{\sqrt{c^2-1}}{3(1+c)} \left( 1+ \frac{1}{1+c} \right), \\ 
\hspace{47mm} (x=1) \\
-\frac{\sqrt{c^2-c^2}}{(1-x^2)(1+c)} - \frac{1}{(x^2-1)^{3/2}} {\rm arccos} \frac{x^2+c}{x(1+c)}, \\ 
\hspace{41mm} (1<x \leqslant c) \\
0 \hspace{44mm} (x>c).
\end{array}
\right.
\end{equation}

The halo's convergence is then, if we assume that all sources are at the same redshift $z_s$,
\begin{equation}
\kappa(x,z_s) = \frac{\Sigma(x)}{\Sigma_{\rm crit}} = 2 \frac{\rho_s r_s}{\Sigma_{\rm crit}} g(x),
\end{equation}
where the critical density is defined as
\begin{equation} \label{eq_sigma_crit}
\Sigma_{\rm crit} = \frac{c^2}{4\pi G} \frac{D_{\rm s}}{D_{\rm d} D_{\rm ds}}.
\end{equation}

In equation (\ref{eq_sigma_crit}), $c$ is the speed of light, $G$ the gravitation constant, and $D_{\rm s}$, $D_{\rm d}$ and $D_{\rm ds}$ are the angular-diameter distances of the source, of the lens, and between the lens and the source, respectively.

To take the redshift distribution $p_z(z_s)$ of source galaxies, we follow \cite{seitz97}, \cite{bartelmann01} and \cite{wk02} and introduce the function
\begin{equation}
Z(z_s;z_d) \equiv \frac{{\rm lim}_{z_s \rightarrow \infty} \Sigma_{\rm crit}(z_d;z_s)}{\Sigma_{\rm crit}(z_d;z_s)} = \frac{\Sigma_{{\rm crit},\infty}}{\Sigma_{\rm crit}(z_d;z_s)},
\end{equation}
where $z_d$ is the lens' redshift, and $\Sigma_{{\rm crit},\infty}$ is the critical density for a source at infinite redshift. This function allows us to project sources with a certain redshift distribution on a single redshift $z_s$ verifying $Z(z_s)=\left< Z \right>$, with $\left< Z \right> = \int \dd z_s p_z(z_s) Z(z_s;z_d)$. One can then link the halo's convergence with a source at $z_s$ to that with infinite redshift source, $\kappa(x,z_s)=\kappa(x) Z(z_s;z_d)$ (\citealt{seitz97}). The convergence given by a source population distributed in redshift along $p_z(z_s)$ is then given by $\kappa(x,p_z(z_s)=\left< Z \right> \kappa(x)$.

The signal-to-noise ratio (Eq. \ref{eq_nu1}), for a NFW halo is then \
\begin{equation} \label{sbnfw2}
\nu_{\rm NFW} = 2 \sqrt{2 \pi} \left< Z \right> \frac{\sqrt{n_g}}{\sigma_\gamma} \frac{\rho_s r_s^2}{D_{\rm d} \Sigma_{\rm crit,\infty}} \sqrt{G(c)}
\end{equation}
where $G(c) = \int_0^c {\rm d}x ~x g(x)^2$ is a function of the halo's concentration only, and is well fitted, for $1 \leqslant c \leqslant 200$ (the maximum deviation being of order 0.5\%), by
\begin{equation}
G(c) \approx \frac{0.131}{c^2} - \frac{0.375}{c} +0.388 - 5 \times 10^{-4} c - 2.8 \times 10^{-7} c^2.
\end{equation}

\end{document}